\UseRawInputEncoding
\documentclass[12pt]{article}
\usepackage[cmtip,arrow]{xy}
\usepackage{pb-diagram,pb-xy}
\usepackage{amsmath}
\usepackage[psamsfonts]{amssymb}
\usepackage{cmmib57}
\usepackage{amsmath,amssymb,amscd}
\usepackage{tikz-cd}

\newcommand{\bPf}{\par\vspace*{-4pt}\indent{\sc Proof.}\enskip}
\newcommand{\ePf}{\medskip}
\def\QED{\hskip0.1em\hfill\null\ \null\nobreak\hfill\kern3pt\vbox{\hrule\hbox
   {\vrule\kern1pt\vbox{\kern1.7pt\hbox{$\scriptscriptstyle{QED}$}
    \kern0.2pt}\kern1pt\vrule}\hrule}}

\def\END{\hskip0.1em\hfill\null\ \null\nobreak\hfill\kern3pt\vbox{\hrule\hbox
   {\vrule\kern1pt\vbox{\kern1.7pt\hbox{$\,\,\,\vspace{5pt}$}
    \kern0.2pt}\kern1pt\vrule}\hrule}}
\newtheorem{theorem}{Theorem}[section]
\newtheorem{lemma}[theorem]{Lemma}
\newtheorem{corollary}[theorem]{Corollary}
\newtheorem{proposition}[theorem]{Proposition}
\newtheorem{remark}[theorem]{Remark}
\newtheorem{definition}[theorem]{Definition}
\newtheorem{example}[theorem]{Example}

\DeclareMathOperator{\byd}{{\raisebox{.1ex}{:}{=}}}
\newcommand{\bCd}{\beq \begin{CD}}
\newcommand{\eCd}{\end{CD}\eEq}
\newcommand{\bcd}{\beq \begin{CD}}
\newcommand{\ecd}{\end{CD}\eeq}
\newcommand{\ben}{\begin{enumerate}}
\newcommand{\een}{\end{enumerate}}
\newcommand{\bEq}{\begin{eqnarray}}
\newcommand{\eEq}{\end{eqnarray}}
\newcommand{\beq}{\begin{eqnarray*}}
\newcommand{\eeq}{\end{eqnarray*}}
\newcommand{\bDf}{\begin{definition}\em}
\newcommand{\eDf}{\end{definition}}
\newcommand{\bLm}{\begin{lemma}}
\newcommand{\eLm}{\end{lemma}}
\newcommand{\bPr}{\begin{proposition}}
\newcommand{\ePr}{\end{proposition}}
\newcommand{\bTh}{\begin{theorem}}
\newcommand{\eTh}{\end{theorem}}
\newcommand{\bCr}{\begin{corollary}}
\newcommand{\eCr}{\end{corollary}}
\newcommand{\bRm}{\begin{remark}\em}
\newcommand{\eRm}{\end{remark}}
\newcommand{\bEx}{\begin{example}\em}
\newcommand{\eEx}{\end{example}}
\newcommand{\C}{\mathbb{C}}

\newcommand{\ie}{{\em i.e$.$} }
\newcommand{\eg}{{\em e.g$.$} }
\newcommand{\R}{I\!\!R}
\newcommand{\A}{\forall}

\newcommand{\mto}{\mapsto}

\newcommand{\der}{\partial}

\newcommand{\balp}{\boldsymbol{\alp}}

\newcommand{\blam}{\boldsymbol{\lam}}

\newcommand{\bnu}{\boldsymbol{\nu}}

\newcommand{\bSigma}{\boldsymbol{\Sigma}}
\newcommand{\cA}{\mathcal{A}}

\newcommand{\cC}{\mathcal{C}}

\newcommand{\cE}{\mathcal{E}}
\newcommand{\cF}{\mathcal{F}}

\newcommand{\cH}{\mathcal{H}}

\newcommand{\cJ}{\mathcal{J}}
\newcommand{\cK}{\mathcal{K}}
\newcommand{\cL}{\mathcal{L}}

\newcommand{\cR}{\mathcal{R}}
\newcommand{\cS}{\mathcal{S}}

\newcommand{\cV}{\mathcal{V}}
\newcommand{\cW}{\mathcal{W}}

\newcommand{\cZ}{\mathcal{Z}}

\newcommand{\bp}{\boldsymbol{p}}
\newcommand{\bq}{\boldsymbol{q}}

\newcommand{\bx}{\boldsymbol{x}}
\newcommand{\by}{\boldsymbol{y}}

\newcommand{\bF}{\boldsymbol{F}}
\newcommand{\bG}{\boldsymbol{G}}
\newcommand{\bH}{\boldsymbol{H}}

\newcommand{\bK}{\boldsymbol{K}}

\newcommand{\bP}{\boldsymbol{P}}
\newcommand{\bQ}{\boldsymbol{Q}}
\newcommand{\bR}{\boldsymbol{R}}

\newcommand{\bW}{\boldsymbol{W}}
\newcommand{\bX}{\boldsymbol{X}}
\newcommand{\bY}{\boldsymbol{Y}}

\newcommand{\sub}{\subset}

\newcommand{\com}{\!\circ\!}
\newcommand{\ten}{\!\otimes\!}

\newcommand{\alp}{\alpha}
\newcommand{\bet}{\beta}
\newcommand{\gam}{\gamma}

\newcommand{\zet}{\zeta}

\newcommand{\lam}{\lambda}
\newcommand{\sig}{\sigma}

\newcommand{\ome}{\omega}

\newcommand{\Lam}{\Lambda}

\newcommand{\For}{{\Lambda}}

\newcommand{\Var}{{\mathcal{V}}}
\newcommand{\Thd}{{\Theta}}

\title{\large{Higgs fields induced by Yang--Mills type Lagrangians on gauge-natural prolongations
of principal bundles
}}

\author{ 
{\normalsize Marcella Palese\footnote{Corresponding Author} \, and Ekkehart Winterroth}
\\ {\footnotesize Department of Mathematics,
University of Torino}
\\
{\footnotesize via C. Alberto 10, 10123 Torino, Italy} 
\\  {\footnotesize e--mail: 
{\sc marcella.palese@unito.it, ekkehart.winterroth@unito.it}}}
\date{}
\begin{document}
\maketitle

\begin{abstract}
We address some new issues concerning spontaneous symmetry breaking.
We define classical Higgs fields for gauge-natural invariant Yang--Mills type Lagrangian field theories through the requirement of the existence of {\em canonical} covariant gauge-natural conserved quantities.
As an illustrative example we consider the `gluon Lagrangian', \ie a Yang--Mills Lagrangian on the $(1,1)$-order gauge-natural bundle of $SU(3)$-principal connections, and canonically define  a `gluon' classical Higgs field through the split reductive structure induced by the kernel of the associated gauge-natural Jacobi morphism. 
\end{abstract}

\noindent {\bf Key words}: Yang-Mills Lagrangian; reduced principal bundle; reduced Lie algebra; classical Higgs field; Cartan connection.

\noindent {\bf 2010 MSC}: 55N30, 53Z05
, 58A20 
, 55R10
, 58A12
, 58E30.

\section{Introduction}

The aim of this paper is to provide the definition of a classical Higgs field {\em canonically induced by the invariance of a gluon Yang-Mills Lagrangian} with respect to the gauge-natural infinitesimal transformations of the bundle of $SU(3)$-connections, seen as a $(1,1)$-order gauge-natural affine bundle; some preliminary results have been sketched in \cite{PaWi16_EPJWC}. 

In a series of previous papers (see, in particular,  \cite{PaWi04,PaWi03,PaWi11}) we have shown that we can suitably resort to {\em Jacobi equations for invariant variational problems} which not only assure stability of critical sections according with a classical approach, see \eg \cite{AtBo83,Bou87}, but in addition, {\em define canonical covariant conserved quantities}. There are also some topological aspects involved; for more information see \cite{PaWi17}.

There is an important point here: the entries of Jacobi equations are not general variations, but {\em vertical parts of gauge-natural lifts}. Note that, in general, these {\em are not} gauge-natural lifts themselves, \ie in general the Lagrangian is not invariant with respect to vertical parts of gauge-natural lifts.

In principle, by this approach, one could obtain principal bundle reductions different from known spontaneous symmetry breaking. Such reductions are strictly related with {\em the requirement of the existence of canonical covariant conserved quantities} associated with gauge-natural invariant Lagrangians by the Noether Theorems, in particular by the Second Noether Theorem.

As an example of application we deal with the gauge-natural Jacobi equations associated with the `gluon' Lagrangian; this enables us to define a {\em canonical classical Higgs field}, that is a canonical reduction of the relevant principal bundle structure. For a gluon Lagrangian within our approach the relevant principal bundle structure is not a $SU(3)$-principal bundle, but its $(1,1)$-order gauge-natural prolongation. 

It is indeed well established that classical physical fields can be described as sections of bundles associated with some gauge-natural prolongations of principal bundles,
by means of suitable left actions of Lie groups on manifolds.
For basics on gauge-natural prolongations and applications in Physics, see \cite{Ec81,KMS93} and \cite{FaFr03}. 
Within our picture  infinitesimal invariant transformations of the Lagrangian will be gauge-natural prolongations of infinitesimal principal automorphisms, lifted to an associated gauge-natural bundle. A gauge-natural Lagrangian is indeed a Lagrangian which is invariant with respect to any of such lifts.

Accordingly, within our approach to symmetry breaking the {\em variation vector fields} are, in fact,  Lie derivatives 
of sections of gauge-natural bundles (\ie of fields)  taken with respect to gauge-natural lifts of infinitesimal 
automorphisms of the underlying principal bundle.  We are inspired by the seminal work by Emmy Noether \cite{Noe18}, who essentially takes as variations vertical parts of generators of infinitesimal invariant transformations of a Lagrangian, see \eg the discussion in  \cite{PaWi17phil}.

Concerning a canonical definition of a Lie derivative of classical physical fields, we formerly tackled the problem how to coherently
define the lift of infinitesimal transformations of the base manifolds up to the bundle
of physical fields, so that
right-invariant infinitesimal automorphisms of the structure bundle would
define the transformation laws of the fields themselves.  
We obtained  an adapted version of the Second Noether Theorem within finite order variational sequences on gauge-natural bundles  
whereby we related the Noether identities to the second variation of a Lagrangian.  We thus
characterized {\em canonical} `strong' (or `off shell') conserved currents through the kernel of a {\em gauge-natural Jacobi morphisms}; for more detail, see \eg  in particular  \cite{PaWi03}, and \cite{FFPW10,PaWi07,PaWi08,PaWi08b}.
 
Indeed, along such a kernel the  gauge-natural lifts of infinitesimal principal 
automorphism are given in terms of the corresponding infinitesimal diffeomorphisms (their projections) 
on the base manifolds in a canonical (although not natural) way.
A canonical determination of Noether conserved quantities is obtained on a reduced sub-bundle of the gauge-natural prolongation of the structure bundle; such a reduction is determined by the invariance properties of a given variational problem (\ie invariant Lagrangian action). 
Connections can be characterized by means of such a canonical reduction 
and conserved quantities can be characterized in terms of Higgs fields on gauge principal bundles presenting the more complex structure of a gauge-natural prolongation, see \cite{FFPW08,FFPW10,PaWi04,PaWi03,PaWi09,PaWi11,PaWi13,PaWi16_EPJWC}.

\section{Variational problems on gauge-natural prolongations modulo contact structures, and lifts}

Let us shortly  summarize the geometric frame and, in particular, some useful concepts of prolongations, mainly with the aim of fixing the notation; for details about (gauge-natural) prolongations see \eg  \cite{Sau89} and \cite{Ec81,KMS93}.

Let $\pi : \bY \to \bX$ be a fibered manifold,
with $\dim \bX = n$ and $\dim \bY = n+m$.
For $s \geq q \geq 0$ integers we deal with the $s$--jet space $J_s\bY$ of equivalent (at a point) classes of
$s$--jet prolongations of (local) sections
of $\pi$ (\ie equivalence classes of local sections such that their partial derivatives from order $0$ up to order $s$ coincide at a fixed point); in particular, we set, with obvious meaning, $J_0\bY \equiv \bY$.  There exist  natural fiberings
$\pi^s_q: J_s\bY \to J_q\bY$, $s \geq q$, $\pi^s: J_s\bY \to \bX$, and,
among these, the {\em affine} fiberings $\pi^{s}_{s-1}$ which defines the contact structure at the order $s$. This structure plays a fundamental r\^ole in the calculus of variations on fibered manifolds.
We denote by $V\bY$ the vector sub-bundle of the tangent
bundle $T\bY$ of vector fields  on $\bY$ which are vertical with respect to the fibering $\pi$.

For $s\geq 1$, taking a slight abuse of notation, we fix a natural splitting induced by the natural contact structure on
{\em finite order} jets bundles (see \eg \cite{Kru90,Sau89})
\beq
J_{s}\bY \times_{J_{s-1}\bY}T^*J_{s-1}\bY =
J_s\bY \times_{J_{s-1}\bY}\left(T^*\bX\oplus V^*J_{s-1}\bY\right)\,.
\eeq

Given a projectable vector field $\Xi : J_{s}\bY \to TJ_{s}\bY$, the above splitting yields $\Xi \, \com \, \pi^{s+1}_{s} = \Xi_{H} + \Xi_{V}$, where
$\Xi_{H}$ and $\Xi_{V}$ are, respectively, the  horizontal and the vertical part of $\Xi$ along $\pi^{s+1}_{s}$ and, if  we have in local adapted coordinates $\Xi = \Xi^{\gam}\der_{\gam} + \Xi^i_{\balp}\der^{\balp}_i$, then we
have $\Xi_{H} = \Xi^{\gam}d_{\gam}$ and
$\Xi_{V} = (\Xi^i_{\balp} - y^i_{\balp + \gam}\Xi^{\gam}) 
\der^{\balp}_{i}$. Here  $d_{\gam}$ is the total derivative (the horizontal lift of $\der_\gamma$ on $J_{s+1}\bY$) and  $\balp$ is a multiindex of lenght $s$.  
As well known, the above splitting induces also a decomposition of the exterior differential
on $\bY$, $(\pi^{r+1}_r)^*\circ d = d_H + d_V$, where $d_H$ and $d_V$ are
called the \emph{horizontal} and \emph{vertical differential}, respectively \cite{Sau89}.
For they are obtained by pull-back on the upper order, such decompositions always rise the order of the objects.

The fibered splitting induced by the contact structure on finite order jets yields a {\em differential forms sheaf splitting} in contact components of different degree, so that a sort of
`horizontalization'  $h$ can be suitable defined as the projection on the summand of lesser contact degree; see \eg \cite{Kru90} and the review in \cite{PaRoWiMu16}.

Now, by an abuse of notation, let us denote by $\ker h$ $+$
 $d\ker h$ the induced sheaf
generated by the presheaf $\ker h$ $+$ $d\ker h$ in the standard way ($d$ is an epimorphism of presheaves, but not of sheaves).
We set $\Thd^{*}_{s}$ $\doteq$ $\ker h$ $+$
$d\ker h$ and $\Var^{*}_s=\For^{*}_s / \Thd^{*}_{s}$. We have {\em the $s$-th order variational sequence}
$0 \to \R_{Y} \to \Var^{*}_{s}$, which is a resolution (by soft sheaves of classes of differential forms) of the constant sheaf $ \R_{Y} $ \cite{Kru90}.
 
The representative of a section $\lam\in\Var^{n}_s$ is a Lagrangian of order 
$(s+1)$ of the standard literature. 
Furthermore $\cE_{n}(\lam) \in \Var^{n+1}_{s}$ is the class of Euler--Lagrange morphism associated with $\lam$. If we let $\gam \in \Var^{n+1}_{s}$,
the class of morphism $\cE_{n+1}(\gam)$ is called the Helmholtz morphism associated with $\gam$; the kernel of its canonical representation reproduces Helmholtz conditions of local
variationality. For details about representations of the variational sequences by differential forms see \cite{PaRoWiMu16} and references therein. 
Within this framework the Jacobi morphism can be characterized, see \cite{PaWi03}, and the more recent \cite{AcPa17} involving the representation by the interior Euler operator.

\subsection{Gauge--natural lift}

If ${\zet}$ is a  suitable representation (see later), in the following we shall consider variational sequences on fibered manifolds $\bY_{\zet}$ which have, in particular, the structure of a  {\em gauge-natural bundle} (see the standard sources \cite{Ec81,KMS93} for gauge-natural bundles and \cite{FFP01} for an approach to variational sequences and conservation laws in this framework).

Denote by $\bP\to\bX$ a principal bundle with structure group $\bG$, $\textstyle{dim}\bX=n$, by $L_{k}(\bX)$ the bundle of $k$--frames 
in $\bX$.
For $r\leq k$ the {\em gauge-natural prolongation of $\bP$}, 
$\bW^{(r,k)}\bP$ $\doteq$ $J_{r}\bP \times_{\bX}L_{k}(\bX)$, is a principal bundle over $\bX$ with structure group the semi-direct product 
$\bW^{(r,k)}_{n}\bG \equiv T^{r}_{n}\bG\rtimes GL_{k}(n)$, with $GL_{k}(n)$  group of $k$--frames 
in $\mathbb{R}^{n}$ while $T^{r}_{n}\bG$ is the space of $(r,n)$-velocities on $\bG$.

Let $\bF$ be a manifold and $\zet: \bW^{(r,k)}_{n}\bG \times_{}\bF\to\bF$ be 
a left action of $\bW^{(r,k)}_{n}\bG$ on $\bF$. 
To the induced right action on $\bW^{(r,k)}\bP\times \bF$ it is associated a {\em gauge-natural bundle} of order 
$(r,k)$ defined by $\bY_{\zet} \doteq \bW^{(r,k)}\bP\times_{\zet}\bF$.

Denote now by $\cA^{(r,k)}$ the sheaf of right invariant vector fields 
on $\bW^{(r,k)}\bP$ (it is a vector bundle over $\bX$). 

\bDf
A {\em gauge-natural lift} is defined as the functorial map 
\beq
\mathfrak{G} : \bY_{\zet}  \times_{\bX} \cA^{(r,k)} \to T\bY_{\zet} \,:
(\by,\bar{\Xi}) \mto \hat{\Xi} (\by) \,
\eeq
 where, for any $\by \in \bY_{\zet}$, one sets: $\hat{\Xi}(\by)=
\frac{d}{dt} [(\Phi_{\zet \,t})(\by)]_{t=0}$,
and $\Phi_{\zet \,t}$ denotes the (local) flow corresponding to the 
gauge-natural lift of $\Phi_{t}$, \ie obtained modulo the representation \cite{Ec81,KMS93}. 
\eDf

The above map lifts any right-invariant local automorphism $(\Phi,\phi)$ of the 
principal bundle $W^{(r,k)}\bP$ into a unique local automorphism 
$(\Phi_{\zet},\phi)$ of the associated bundle $\bY_{\zet}$. 
This lifting depends linearly on derivatives up to order $r$ and $k$, respectively, of the components  $\xi^{A}$ and 
$\xi^{\mu}$  of the corresponding infinitesimal automorphism of $\bP$ .
Its infinitesimal version associates to any projectable $\bar{\Xi} \in \cA^{(r,k)}$, a unique {\em projectable}  (over the same tangent vector field on the base manifold) vector field 
$\hat{\Xi} \byd \mathfrak{G} (\bar{\Xi})$ on $\bY_{\zet}$.
Such a functor defines a class of parametrized contact transformations.

This map fulfils the following properties (see \cite{KMS93}):
 $\mathfrak{G}$ is linear over $id_{\bY_{\zet}}$;
we have $T\pi_{\zet}\circ\mathfrak{G} = id_{T\bX}\circ 
\bar{\pi}^{(r,k)}$, 
where $\bar{\pi}^{(r,k)}$ is the natural projection
$\bY_{\zet} \times_{\bX} 
\cA^{(r,k)} \to T\bX$;
 for any pair $(\bar{\Lam},\bar{\Xi})$ $\in$
$\cA^{(r,k)}$, we have
$\mathfrak{G}([\bar{\Lam},\bar{\Xi}]) = [\mathfrak{G}(\bar{\Lam}), \mathfrak{G}(\bar{\Xi})]$.

We have the coordinate expression of $\mathfrak{G}$
\beq
\mathfrak{G} = d^\mu \ten \der_\mu + d^{A}_{\bnu}
\ten (\cZ^{i\bnu}_{A} \der_{i}) + d^{\nu}_{\blam}
\ten (\cZ^{i\blam}_{\nu} \der_{i}) \,,
\eeq
with $0<|\bnu|<k$, $1<|\blam|<r$ and 
$\cZ^{i\bnu}_{A}$, $\cZ^{i\blam}_{\nu}$ $\in C^{\infty}(\bY_{\zet})$ 
are suitable functions which depend only 
on the fibers of the bundle.

\subsection{Variations: Lie derivative of sections and vertical parts of gauge-natural lifts}

When deriving Euler--Lagrange field equations it is of fundamental importance to be able to say something on how their solutions behave under the action of infinitesimal transformations (automorphisms) of the gauge-natural bundle. The geometric object providing us with such an information is, of course, the Lie derivative. Let $\gam$ be a (local) section of $\bY_{\zet}$, $\bar{\Xi}$ 
$\in \cA^{(r,k)}$ and let us denote $\hat\Xi
\doteq \mathfrak{G}(\bar{\Xi})$ its gauge-natural lift. 
Following \cite{KMS93} we
define the {\em 
generalized Lie derivative} of $\gam$ along the projectable vector field 
$\hat{\Xi}$ to be the (local) section $\pounds_{\bar{\Xi}} \gam : \bX \to V\bY_{\zet}$, 
given by ($\xi$ is the projection vector field on the base manifold)
\beq
\pounds_{\bar{\Xi}} \gam = T\gam \circ \xi - \hat{\Xi} \circ \gam \,.
\eeq
Due to the functorial nature of $\hat{\Xi}$, the Lie derivative operator acting on sections of gauge-natural 
bundles inherits some useful linearity properties and, in particular, it is an homomorphism of Lie algebras. In the view of Noether's theorems, the interest of the Lie derivative of sections is due to the fact that it is possible to relate it with the vertical part of a gauge-natural lift, \ie for any gauge-natural lift, we have that 
\beq
\hat{\Xi}_V = - \pounds_{\bar{\Xi}} \,.
\eeq
Inspired by Noether, we shall restrict allowed variations to vertical parts of gauge-natural lifts.

\section{Variationally featured classical `gluon' Higgs fields}

As well known the Standard Model is a gauge theory with structure group $\bG = SU(3)\times SU(2) \times U(1)$. One can consider the coupling with gravity by adding the principal spin bundle $\bar{\Sigma}$ with structure group Spin$(1,3)$; the structure bundle of the whole theory can be then taken to be the fibered product $\bSigma= \bar{\Sigma} \times_{\bX} \bP$.
There is an action of Spin$(1,3)$ on a spinor matter manifold $V=\C ^k$ and therefore a representation Spin$(1,3)\times SU(3)\times SU(2) \times U(1)\times V
$, given by a choice of Dirac matrices for each component of the spinor field.
A corresponding Lagrangian is therefore given by
$\lam = \bar{\psi}(i\gam_\mu D^\mu -m)\psi - \frac{1}{4}(\cF_{\mu\nu} \cF^{\mu\nu} + \cF^A_{\mu\nu} \cF_A^{\mu\nu} +\cF^a_{\mu\nu} \cF_a^{\mu\nu})$.

Experimental evidence concerned with symmetry properties of fundamental interactions shows the phenomenon of  {\em  spontaneous symmetry breaking} suggesting the presence of a scalar field called the Higgs boson on which the spin group acts trivially. A clear introduction to those topics can be found, \eg in 
 \cite{Peccei00}.

For an illustrative purpose, let us then restrict to pure gluon fields assumed to be critical sections of the `gluon Lagrangian' 
$\lam_{gluon}= - \frac{1}{4}\cF^a_{\mu\nu} \cF_a^{\mu\nu}$.

In this note, we shall therefore restrict to a principal bundle $\bSigma$ with structure group $\bG =  SU(3)$, such that $\bSigma / SU(3)= \bX$ and $\textstyle{dim}\bX =4$.

Recall that $W_{4}^{(1,1)} \bG$ is the semi-direct product of $GL(4, \mathbb{R})$ 
on $T^{1}_{4}\bG$, where $GL(4,\mathbb{R})$  is the structure group of linear frames in $\mathbb{R}^{4}$.

The set $\{j_0^k\alp : \alp: \mathbb{R}^4 \to \mathbb{R}^4\}$, with $\alp(0) = 0$  locally invertible,
equipped with the jet composition $j_0^k\alp\circ j_0^k\alp' := j_0^k(\alp \circ \alp')$ is a Lie group called
the $k$-th differential group and denoted by $G^k_4$.
For $k = 1$ we have, of course, the identification $G^1_4 \simeq GL(4, \mathbb{R})$.
The principal bundle over $\bX$ with group $G^k_4$ is called the $k$-th order frame bundle over $\bX$ , $L_k(\bX)$.
For $k = 1$ we have the identification $L_1(\bX) \simeq L\bX$, where $L\bX$ is the usual bundle of linear frames over $\bX$.
 
Unlike $J_1\bSigma$, $W^{(1,1)} \bSigma$ is a principal bundle over $\bX$ with structure group  
\beq
W^{(1,1)}_4\bG \doteq  T^1_4 SU(3)  \rtimes   GL(4,\mathbb{R}) \,
\eeq
 $T^1_4 SU(3)$ being  the Lie group  of $(4,1)$-velocities of $SU(3)$ (if $u: \mathbb{R}^4 \to SU(3)$, a generic element of  $j^1_0 u \in T^1_4 SU(3)$ is represented by 
$g^b=u^b (0)$ and $g^b_\nu=(\der_\nu (g^{-1}\cdot u(x))|_{x=0})^b$).
The group multiplication on $W^{(1,1)}\bG$  being  
\beq
(j_0^1\alp, j_0^1a) \odot (j_0^1 \bet, j_0^1 b) \doteq ( j_0^1(\alp\circ\bet), j_0^1( (a \circ\bet)\cdot b ))
\eeq and denoting  by  $\cdot_r$ the right action of $SU(3)$ on $\bSigma$, the right action of $W^{(1,1)}_4\bG$ on $W^{(1,1)}\bSigma$
is then defined by 
\beq
(j_0^1\rho, j_x^1\sig) \odot (j_0^1\alp, j_0^1 a) \doteq (j_0^1(\rho\circ\alp), j_x^1 (\sig\cdot_r (a \circ \alp^{-1} \circ \rho^{-1})))\,. 
\eeq

\bRm 
It is known that the bundle of principal connections on $\bSigma$ is a gauge-natural bundle associated with the gauge-natural prolongation 
$W^{(1,1)}\bSigma$. 
Indeed, consider the action $\zeta$ induced by the adjoint representation:
\beq
\zeta &: &W_{4}^{1,1} \bG \times (\mathbb{R}^{4})^{*} \ten \, \mathfrak{su}(3)  \to (\mathbb{R}^{4})^{*} \ten \, \mathfrak{su}(3)\\
&:& ( (g^b  , g^c_{\mu} , \alp^\sig_\rho) , f^a_\nu )\mto (Ad_g)^a_b ( f^b_\sig  - g^b_\sig ) \bar{\alp}^\sig_\nu\,,
\eeq
where $(Ad_g)^a_b$ are  the coordinate expression of the adjoint representation of $\bG=SU(3)$ and  $g^b  , g^c_{\mu} $ denote natural coordinates on $T^1_4 SU(3)$.
The sections of the associated bundle 
\beq
\cC(\bSigma)\doteq W^{(1,1)}\bSigma \times_{\zeta} (\mathbb{R}^{4})^{*} \ten \,\mathfrak{su}(3)\to \bX
\eeq
are in  $1$ to $1$ correspondence with the principal connections on $\bSigma$ and are called $SU(3)$-connections. Clearly, by construction, $\cC(\bSigma)$ is a $(1,1)$-order  gauge-natural affine bundle; see \eg \cite{KMS93} and \cite{FaFr03} for some details, especially presentations in local coordinates, and applications in Physics.
\eRm 

Note that the Lie algebra of $W^{(1,1)}_4 SU(3)$ is the semi-direct product of $\mathfrak{gl}(4,\mathbb{R})$ with the Lie algebra, $\mathfrak{t}^1_4\mathfrak{su}(3)$,  of $T^1_4 SU(3)$.
It is easy to characterize the semi-direct product of the two Lie algebras, from now on denoted by     
$\cS$, as the direct sum $\mathfrak{t}^1_4\mathfrak{su}(3) \oplus \mathfrak{gl}(4,\mathbb{R})$  with a bracket induced by the right action of $GL(4,\mathbb{R})$ on $T^1_4 SU(3)$ given by the jet composition, in particular by the induced Lie algebra homomorphism $\mathfrak{t}^1_4\mathfrak{su}(3)  \to \textstyle{hom} (\mathfrak{gl}(4,\mathbb{R}))$; given a base of $\mathfrak{t}^1_4\mathfrak{su}(3)  \rtimes \mathfrak{gl}(4,\mathbb{R})$; the adjoint representation of the Lie group $W_{4}^{(1,1)} SU(3)$ is also readily defined (see \eg \cite{JaVo09},  and \cite{Von10}  $\S 1.3$). 

Local coordinates on $W_{4}^{(1,1)} SU(3)$ are given by $(g^b, g^b_\sig;  \alp^\mu_\sig)$, and let us denote  the induced local coordinates on $\cS$ by $(Y^a, Y^a_\mu, X^\mu_\sig)$.
Local generators of the tangent space are of course partial derivative with respect to such local coordinates.

Consider the right action $R_{\hat{g}} : W^{(1,1)}\bSigma \to W^{(1,1)}\bSigma$, $\hat{g}\in W^{(1,1)}_4 SU(3)$.
Let
$\Xi$ be a right invariant vector field on $W^{(1,1)}\bSigma$. In coordinates we have
$ \Xi=\xi^\lam\der_\lam + \Xi^A
 \tilde{\mathfrak{b}}_A$
where $(\tilde{\mathfrak{b}}_A)$ is the base of vertical right invariant vector fields on $W^{(1,1)}\bSigma$ which are induced by the base $(\mathfrak{b}_A)$ of $\cS$ (here the index $A$ encompasses all indices in the Lie algebra $\cS$). 
They are sections of the bundle $TW^{(1,1)}\bSigma/W^{(1,1)}_4 SU(3)\to \bX$. We have $ \tilde{\mathfrak{b}}_A = (R_{\hat{g}}  )^B_A  \der_B $, where the invertible matrix $(R_{\hat{g}}  )^B_A$ is the matrix representation of $TR_{\hat{g}}$.
It is clear that so-called Gell-Mann matrices $\lam_a$ are matrix representations of $\mathfrak{b}_a$
and they therefore induce $\tilde{\mathfrak{b}}_a$ in the standard way.  Analogously a matrix representation can be obtained for $\mathfrak{b}^\mu_a$, and $\mathfrak{b}_\mu^a$,  being essentially 
\beq
T^1_4 SU(3)\rtimes GL(4, \mathbb{R})\simeq (SU(3) \times (\mathbb{R}^{4})^{*} \ten \, \mathfrak{su}(3)) \rtimes  GL(4, \mathbb{R})\,.
\eeq

\subsection{Split reductive structure induced by gauge-natural invariant `gluon' Lagrangians}

The linearity properties of the gauge-natural lift $\hat{\Xi}$ of  infinitesimal automorphisms of $W^{(1,1)}\bSigma$ to the bundle $\cC(\bSigma)$ of $SU(3)$-connections (see \eg \cite{FaFr03} for the coordinate expressions)
enable to suitable define a {\em gauge-natural generalized Jacobi 
morphism} associated with a Lagrangian $\lam$ and the {\em variation vector field $\hat{\Xi}_{V}$}, the vertical part of  $\hat{\Xi}$,
\ie the bilinear morphism 
\beq
\cJ(\lam_{gluon},\hat{\Xi}_{V}) \doteq 
\hat{\Xi}_{V}\rfloor {\bar\cE} (\hat{\Xi}_{V}\rfloor\cE (\lam_{gluon}))\,,
\eeq
 where $\cE$ is the Euler--Lagrange morphism on the jet space of $\bY\equiv \cC(\bSigma)$, while ${\bar\cE}$ is the Euler--Lagrange morphism on the space extended by the components of $\hat{\Xi}_{V}$ \cite{PaWi03,PaWi07}. 
 
Gauge-natural lifts of infinitesimal principal automorphisms the vertical part of which are in the kernel $\mathfrak{K}\doteq \ker\cJ(\lam_{gluon},\hat{\Xi}_{V})$ are called {\em generalized gauge-natural Jacobi vector fields} and generate {\em canonical} covariant conserved quantities \cite{PaWi04,PaWi03,PaWi08b}. They have the property that the Lie derivative of critical sections are still critical sections, \ie their flow leave invariant the equations  {\em  and} the set of critical sections (although in general they could be not symmetries of the Lagrangian). Such a kernel is a sub-algebra of the Lie algebra of vertical tangent vector field; from a theoretical physics point of view it can be interpreted as an internal symmetry algebra (see later). 
An explicit description of  $\mathfrak{K}$ for $\lam_{gluon}$ is obtained from the equation $\cJ=0$, by inserting the corresponding Euler--Lagrange expressions and the vertical parts of gauge-natural lifts.

We first recall that, in a general context, the kernel of the gauge-natural Jacobi morphism associated with a gauge-natural invariant Lagrangian determines a split reductive structure \cite{PaWi08}. 
\bTh
The kernel $\mathfrak{K}$ defines a canonical split reductive structure on $W^{(r+4s,k+4s)}\bP$.
\eTh

\bPf 
Let $\mathfrak{h}$ be the Lie algebra of right-invariant vertical vector fields on $W^{(r+4s, k+4s)}\bP$ and $\mathfrak{k}$ the algebra of generalized Jacobi vector fields.
It is well known that the Jacobi morphism is self-adjoint along critical sections (it was proved in \cite{GoSt73} for first order field theories and in \cite{AcPa17} for higher order field; this property has been also proved to hold true along any section modulo divergences \cite{FPV05} and within the variational sequence on the vertical bundle of the relevant fibered manifold \cite{PaWi07}).
Therefore we have that $\textstyle{dim}\mathfrak{K}=\textstyle{dim}\textstyle{Coker}\cJ$. If we further consider that $\mathfrak{K}$ is of constant rank \cite{PaWi07} (and thus $\mathfrak{k}$ is a Lie sub-algebra), we get a split structure on $\mathfrak{h}$, given by $\mathfrak{k}\oplus \textstyle{Im}\cJ$.

It is easy to see that the Lie derivative with respect to vertical parts of the commutator between the gauge-natural lift of a Jacobi vector field and (the vertical part of) a lift not lying in $\mathfrak{K}$  {\em 
is not} a solution of Euler--Lagrange equations.  Thus, we have the reductive property $[ \mathfrak{k},\textstyle{Im}\cJ ]=\textstyle{Im}\cJ$ \cite{PaWi03,PaWi07,PaWi08b}.
\ePf

Since the action is effective, the Lie algebra of fundamental vector fields (right-invariant vertical vector fields on $W^{(r+4s, k+4s)}\bP$) and the corresponding Lie sub-algebra (Jacobi right-invariant vertical vector fields on $W^{(r+4s, k+4s)}\bP$) are isomorphic to the corresponding Lie algebras of the Lie groups of the respective principal bundles.

\subsection{Canonical reduction of $\bW^{(1,1)}\bSigma$}

We remark that in the case of an $SU(3)$-connection, the canonical reductive structure is defined on each fiber of  $VW^{(1,1)}\bSigma /  W^{(1,1)}_{4}SU(3)$.
Denote then $\cS\doteq\mathfrak{h}$, $\cR\doteq\mathfrak{k}$ and $\cV\doteq \textstyle{Im}\cJ$; 
by the theorem above, we have a reductive Lie algebra decomposition $\cS\doteq \mathfrak{t}^1_4\mathfrak{su}(3)  \rtimes \mathfrak{gl}(4,\mathbb{R})=\cR\oplus\cV$, with $[\cR,\cV]=\cV$,
where $\cS$ is the Lie algebra of the  structure Lie group $W^{(1,1)}_{4} SU(3)$. 
Note that there exists an isomorphism between $\cV\doteq \textstyle{Im}\cJ_{\bp}$ and $T_{\bx}\bX$ so that $\cV$ turns out to be the image of an horizontal subspace. 
In the case of a $W^{(1,1)}_{4} SU(3)$ gauge-natural bundle, let us denote by $\bR$ the Lie group of the Lie sub-algebra $\mathfrak{k}$. 
As we show in the following, we get a reduction of the principal bundle $W^{(1,1)}_{4} SU(3)$.

Indeed, in the following we state the existence of a principal bundle $\bH\to\bX$, 
where 
$\bR$, the Lie group of the Lie algebra $\cR$, is a closed subgroup of $W^{(1,1)}_{4} SU(3)$.
The principal sub-bundle $\bH\sub \bW^{(1,1)}\bSigma$ is then a
 {\em reduced principal bundle}. The Lie algebra $\cR$ is a reductive Lie sub-algebra of $\mathfrak{t}^1_4\mathfrak{su}(3)  \rtimes \mathfrak{gl}(4,\mathbb{R})\simeq (\mathfrak{su}(3) \ltimes (\mathbb{R}^{4})^{*} \ten \, \mathfrak{su}(3)) \rtimes  \mathfrak{gl}(4,\mathbb{R}) 
\simeq \mathfrak{su}(3) \oplus ( (\mathbb{R}^{4})^{*} \ten \, \mathfrak{su}(3) \rtimes  \mathfrak{gl}(4,\mathbb{R})  )
\simeq (\mathfrak{su}(3) \oplus  \mathfrak{gl}(4,\mathbb{R} ) \oplus ( ( \mathbb{R}^{4})^{*} \ten \, \mathfrak{su}(3) )$.
Such a split reductive structure thus `generates' a canonical (although not natural), {\em variationally induced},  breaking of the symmetry group $W^{(1,1)}_{4} SU(3)$, \ie generates classical Higgs fields in the sense defined later on.

The (gauge-natural) Jacobi fields are (generated by) a Lie sub-algebra of fundamental vector fields on $W^{(1,1)}_{4} SU(3)$; the crucial point here is indeed to characterize such a Lie sub-algebra. 
\subsection{Split reductive structures and Higgs fields in the case of $SU(3)$-connections}

Let us rephrase the above result for our specific case of study.

We have the composite fiber bundle (see \cite{FFPW10,PaWi09}) 
\beq
W^{(1,1)}\bSigma\to W^{(1,1)}\bSigma/\bR\to\bX \,,
\eeq
 where $W^{(1,1)}\bSigma/\bR =W^{(1,1)}\bSigma\times_{W^{(1,1)}_{4} SU(3)} W^{(1,1)}_{4} SU(3)/\bR \to\bX$ is a gauge-natural bundle functorially associated with $W^{(1,1)}\bSigma\times W^{(1,1)}_{4} SU(3)/\bR \to\bX$  by the right action of $W^{(1,1)}_{4} SU(3)$.

The left action of \,$W^{(1,1)}_{4} SU(3)$ on $W^{(1,1)}_{4} SU(3)/\bR$ is defined by  the reductive Lie algebra decomposition. 

\bDf 
According to \cite{Sarda06,Sarda14}, we call a global section $h: \bX\to W^{(1,1)}\bSigma/\bR$ a {\em classical gluon Higgs field}. 
\eDf

A global section $h$ of $W^{(1,1)}\bSigma /\bR\to \bX$  defines a vertical covariant differential and therefore the Lie derivative of fields is constrained and it is parametrized by gluon Higgs fields $h$ characterized by $\mathfrak{K}$ \cite{PaWi11,PaWi13}.

 \subsection{Higgs fields as Cartan connections}\label{Cartan}
Turning back to the case of a generic principal bundle $\bP$, once we have solutions of the Jacobi equations we would like to characterize them as the fundamental vector fields of a reduced principal sub-bundle of $\bP$, which we shall denote by $\bQ$. We can then obtain the Lie sub-algebra as the Lie algebra of invariant vectors produced by the vertical parallelism
of a principal connection on $\bQ$ (see in particular \cite{AlMi95}). 

In other words, we should be able to recognize that  the Jacobi equations select among vertical parts of gauge-natural lifts those vector fields which reproduce invariant tangent vectors on the reduced Lie group.
To do this we have to know or recognize the action of the Lie sub-group of $\bQ$. This action emerges from the structure of split reductive decomposition.

Let now $\textstyle{rank \, ker}\cJ = dim \bX$. It is noteworthy that a specific kind of Cartan connection is defined by the intrinsic structure of an {\em invariant} Lagrangian theory by means of the kernel of the Jacobi morphism.
For a characterization of the bundle of Cartan connections as a gauge-natural bundle, see \cite{Panak}.
 
The following is a general result for invariant Lagrangian theories on gauge-natural bundles; see also \cite{PaWi09}. 
\bPr \label{rank}
Let $\textstyle{rank \, ker}\cJ = dim \bX$. 
Let $\cW$ be the Lie algebra of the Lie group of the principal bundle $W^{(r,k)}\bP$.
A principal Cartan connection is canonically defined by gauge-natural invariant variational problems of finite order.
\ePr

\bPf
Since $\mathfrak{K}$ is a vector sub-bundle of $\cA^{(r,k)}=T\bW^{(r,k)}\bP/\bW_{n}^{(r,k)}\bG$ there exists a principal sub-bundle $\bQ\sub \bW^{(r,k)}\bP$ such that $dim\bQ=dim\cW$,  $\cK = T\bQ/\bK |_{\bq}$, where $\bK$ is the (reduced) Lie group of the Lie algebra $\cK$ and the embedding $\bQ\to  \bW^{(r,k)}\bP$ is a principal bundle homomorphism over the injective group homomorphism $\bK \to \bW_{n}^{(r,k)}\bG$.

Now, if $\ome$ is a principal connection on $\bW^{(r,k)}\bP$, the restriction $\ome |_{\bQ}$ is a Cartan connection of the principal bundle $\bQ\to\bX$. 
In fact, let us consider a principal connection $\bar{\ome}$ on the principal bundle $\bQ$ \ie a $\cK$-invariant horizontal distribution defining the vertical parallelism $\bar{\ome}: V\bQ\to \cK$ by means of the fundamental vector field mapping in the usual and standard way. Since $\cK$ is a sub-algebra of the Lie algebra $\cW$ and $dim\bQ=dim\cW$, it is defined a principal Cartan connection of type $\cW/\cK$, that is a $\cW$-valued absolute parallelism $\hat{\ome}: T\bQ\to \cW$ which is an homomorphism of of Lie algebras, when restricted to $\cK$, preserving Lie brackets if one of the arguments is in $\cK$, and such that $\bar{\ome}=\hat{\ome} |_{V\bQ}$, that means that $\hat{\ome}$ is an extension of the natural vertical parallelism. 

Such a connection  $\hat{\ome}$ is defined as the restriction of the natural vertical parallelism defined by a principal connection $\ome$ on $W^{(r,k)}\bP$ by means of the fundamental vector field mapping $\ome:VW^{(r,k)}\bP\to \cW$ to $T\bQ$. This restriction is, in particular, $\cK$-invariant since is by construction $\cW$-invariant. 

The definition is well done since $T\bQ\sub VW^{(r,k)}\bP$ holds true as a consequence of the split reductive structure on $W^{(r,k)}\bP$. In particular, $\A \bq\in\bQ$, we have  $T_{\bq}\bQ\cap \cH_{\bq}= 0$, where $\cH_{\bq}$, $\A \bp\in \bW^{(r,k)}\bP$ is defined by $\ome$ as 
$T_{\bp}\bW^{(r,k)}\bP=V_{\bp}\bW^{(r,k)}\bP\oplus\cH_{\bp}$; furthermore, $\textstyle{dim} \bX= \textstyle{dim} \cW/\cK$ \cite{Sha97}.
\ePf

\bEx {\em 
Let a Lagrangian theory on a $SU(3)$-principal bundle $\Sigma$ satisfies  the condition $\textstyle{rank \, ker}\cJ = dim \bX$. Let then $\ome$ denotes a principal connection on $W^{(1,1)}\bSigma$; 
$\bar{\ome}$  principal connection on the reduced principal bundle $\bH$
defines the splitting $T_{\bp}\bH\simeq_{\bar{\ome}} \cR\oplus \bar{\cH}_{\bp}$, $\bp\in\bH$. Note that, for each $\bq \in W^{(1,1)}\bSigma$, $T_{\bq} W^{(1,1)}\bSigma \simeq_{\ome} V_{\bq} W^{(1,1)}\bSigma \oplus \cH_{\bq}$.
We find that $V_{\bq} W^{(1,1)}\bSigma \simeq  T_{\bq}\bH \simeq_{\bar{\ome}}  \cR \oplus \bar{\cH}_{\bq}$, $\bq\in\bH$, \ie
 Cartan connection $\hat{\ome} $ of type $\cS/\cR$ is defined, such that $\hat{\ome} |_{V\bH}=\bar{\ome}$  \cite{PaWi09}. 
It  is a connection on $W^{(1,1)}\bSigma=\bH \times_{\bR}W^{(1,1)}_{4} SU(3)\to \bX$, thus a Cartan connection on $\bH\to\bX$ with values in $\cS$, the Lie algebra of the gauge-natural structure group of the theory; it splits into the $\cR$-component which is a principal connection form on the $\cR$-manifold $\bH$, and
the $\cV$-component which is a displacement form; see  \cite{AlMi95} for the geometric frame and for the terminology. 
A gauge-natural Higgs field is therefore a global section of the Cartan horizontal bundle $\hat{\cH}_p$, with $\bp\in \bH$,
it is related with the displacement form defined by  the $\cV$-component of the Cartan connection $\hat{\ome}$ above.
The case of Yang--Mills theories satisfying the rank assumption of Proposition \ref{rank} will be the object of separate researches.
}
\eEx

\subsection{An application to Yang--Mills type Lagrangians on a Minkowskian background}

As for a manageable example of application, let us consider Yang--Mills theories on a Minkowskian background, \ie the {\em space-time manifold} is equipped with a fixed {\em Minkowskian metric} (\ie assume we can choose a system of coordinates in which the metric is expressed in the {\em diagonal form} $\eta_{\mu \nu}$); for details about this example, see \cite{AcPa17}.

Note that, as we shall see, in the case of a `gluon' Lagrangian on a Minkowskian background, the the rank assumption of Proposition \ref{rank} is not satisfied; however, although a Cartan connection cannot be given in this case, we still get a principal bundle reduction. 
Indeed, in the specific case of study, if we would have $\textstyle{rank \, ker}\cJ = dim \bX$ the corresponding Jacobi equations would not admit non zero solutions, \ie we could not construct a Cartan connection because $\textstyle{ker}\cJ$ would be trivial. When  $\textstyle{rank \, ker}\cJ < dim \bX$ (in our example this corresponds to some feature of the curvature) the Jacobi equations admit non zero solutions and principal bundle reductions are obtained. 

In the following it is assumed that the structure bundle of the theory has a {\em semi-simple structure group} $G$.
In this example, lower Greek indices label space-time coordinates, while capital Latin indices label the Lie algebra $\mathfrak{g}$ of $G$. Then, on the bundle of principal connections, introduce coordinates $(x^\mu, \omega^A_\sigma)$.
Consider the {\em Cartan-Killing metric} $\delta$ on the Lie algebra $\mathfrak{g}$, and choose a $\delta$-orthonormal basis $T_A$ in $\mathfrak{g}$; the components of $\delta$ will be denoted $\delta_{AB}$ they raise and lower Latin indices; by $c^D_{EF}$ we denote the structure constants of the Lie algebra.
Let  
\beq 
\Xi=\Xi^Z_\sigma (x^\mu, \omega^A_\sigma)\frac{\partial}{\partial \omega^Z_\sigma} \,,
\eeq 
be a vertical vector field on the bundle of connections.
On  the bundle of vertical vector fields over the bundle of connections, an induced connection (recall that a Minkowskian background is assumed) is defined by
\beq 
\tilde{\Omega} = dx^\mu\otimes (\frac{\partial}{\partial x^\mu}-\omega^B_{\sigma\mu}(x,\phi)\frac{\partial}{\partial\omega_{\sigma}^B}) = dx^\mu\otimes \nabla_\mu \,.
\eeq 
For any pair $(\nu,B)$, the Jacobi equation for the Yang-Mills Lagrangian can be suitably written as
\beq
\eta^{ \nu\sigma }\eta^{\beta \alpha }\left\{
\nabla_\beta\left[\left(\nabla_\alpha\Xi^A_\sigma-\nabla_\sigma\Xi^A_\alpha\right)\delta_{BA}\right]+
F^D_{\beta\sigma}\delta_{AD}c^A_{BZ}\Xi^Z_\alpha\right\}=0 \,,
\eeq
 (this  result was obtained in \cite{AcPa17}).

Let us work out the meaning of these Jacobi equations. Note now that, due to the antisymmetry of $F^D_{\beta\sigma}$ in the lower indices, these equations split in the antisymmetric and symmetric parts
\beq
\eta^{ \nu[\sigma }\eta^{\beta] \alpha }\left\{
\nabla_\beta\left[\left(\nabla_\alpha\Xi^A_\sigma-\nabla_\sigma\Xi^A_\alpha\right)\delta_{BA}\right]+
F^D_{\beta\sigma}\delta_{AD}c^A_{BZ}\Xi^Z_\alpha\right\}=0 \,,
\eeq
and
\beq
\eta^{ \nu(\sigma }\eta^{\beta) \alpha }\left\{
\nabla_\beta\left[\left(\nabla_\alpha\Xi^A_\sigma-\nabla_\sigma\Xi^A_\alpha\right)\delta_{BA}\right] 
\right\}=0 \,.
\eeq

On the other hand, on a Minkowskian background as defined above, $\eta^{\beta \alpha}= 0$ when $\alpha  \neq \beta $, therefore the only non zero terms are given for $\alpha = \beta$, in which case the second equation turns out to be an identity, while the first one gives us the following algebraic constraints
\beq
\eta^{ \nu[\sigma }\eta^{\beta] \alpha }\left\{
F^D_{\beta\sigma}c_{DBZ}\Xi^Z_\alpha
\right\}=0 \,,
\eeq
for each $\nu=\sigma$ and $\alpha = \beta$ and for each $B$.

In particular multypling for $\mathfrak{b}_B$ and summing up, we get
\beq
\eta^{ \nu[\sigma }\eta^{\beta] \alpha }\left\{
F^D_{\beta\sigma}[\mathfrak{b}_D, \mathfrak{b}_Z] \Xi^Z_\alpha
\right\}=0 \,,
\eeq
for each $\nu=\sigma$ and $\alpha = \beta$, \ie
\beq
\eta^{ 0[ 0 }\eta^{\beta] \alpha }\left\{
F^D_{\beta 0}[\mathfrak{b}_D, \mathfrak{b}_Z] \Xi^Z_\alpha
\right\} = 0
\\
\eta^{ 1[ 1 }\eta^{\beta] \alpha }\left\{
F^D_{\beta 1}[\mathfrak{b}_D, \mathfrak{b}_Z] \Xi^Z_\alpha
\right\} = 0
\\
\eta^{ 2[ 2 }\eta^{\beta] \alpha }\left\{
F^D_{\beta 2}[\mathfrak{b}_D, \mathfrak{b}_Z] \Xi^Z_\alpha
\right\} = 0
\\
\eta^{ 3[ 3}\eta^{\beta] \alpha }\left\{
F^D_{\beta 3}[\mathfrak{b}_D, \mathfrak{b}_Z] \Xi^Z_\alpha
\right\}  = 0 \,,
\eeq
which give us
\beq
-
F^D_{1 0}[\mathfrak{b}_D, \mathfrak{b}_Z] \Xi^Z_1
-
F^D_{2 0}[\mathfrak{b}_D, \mathfrak{b}_Z] \Xi^Z_2
-
F^D_{3 0}[\mathfrak{b}_D, \mathfrak{b}_Z] \Xi^Z_3
  = 0 \,,
\eeq

\beq
-
F^D_{0 1}[\mathfrak{b}_D, \mathfrak{b}_Z] \Xi^Z_0
 + 
F^D_{3 1}[\mathfrak{b}_D, \mathfrak{b}_Z] \Xi^Z_3
 +
F^D_{2 1}[\mathfrak{b}_D, \mathfrak{b}_Z] \Xi^Z_2
 = 0 \,,
\eeq

\beq
-
F^D_{0 2}[\mathfrak{b}_D, \mathfrak{b}_Z] \Xi^Z_0
 + 
F^D_{1 2}[\mathfrak{b}_D, \mathfrak{b}_Z] \Xi^Z_1
+ 
F^D_{3 2}[\mathfrak{b}_D, \mathfrak{b}_Z] \Xi^Z_3
 = 0 \,,
\eeq

\beq
-
F^D_{0 3}[\mathfrak{b}_D, \mathfrak{b}_Z] \Xi^Z_0
  +
F^D_{1 3}[\mathfrak{b}_D, \mathfrak{b}_Z] \Xi^Z_1
+
F^D_{2 3}[\mathfrak{b}_D, \mathfrak{b}_Z] \Xi^Z_2
 = 0 \,.
\eeq
In general, we get constraints on the components $\Xi^Z_\mu$ of vertical vector fields lying in the kernel of the Jacobi morphism.

As a first example of application, when non zero solutions exist, it is easy to check that if $\bG=SU(2) \times U(1)$, by inserting the Lie brackets of the corresponding Lie algebra the above equations reduce to a set of three identical equations for each $Z=1,2,3 = \textstyle{dim} \, \, SU(2)$, given by
$\tilde{F}_{\alpha\beta}\Xi^Z_\alpha= 0$, where $\tilde{F}_{\alpha\beta}=F^1_{\alpha\beta}=F^2_{\alpha\beta}=F^3_{\alpha\beta}$, while the presence of null brackets of the generator of $U(1)$ with generators of $SU(2)$ leave $\Xi^4_\alpha$ free. We get an underdetermined system (made of only one equation) for $\Xi^Z_\alpha$, for $Z=1,2,3$, from which, considering $\Xi^Z_\alpha$ as gauge natural lifts, and taking into account the Lie algebra brackets relations, we get $\mathfrak{b}_1=\mathfrak{b}_2=\mathfrak{b}_3=0$, while $\mathfrak{b}_4$ remains free.
We have therefore a reduction of $SU(2) \times U(1)$ to $U(1)$ (similarly to spontaneous symmetry breaking).

\medskip 

Let us now come back to the case of  $SU(3)$-connections. Working out the Jacobi equations with the $\mathfrak{su}(3)$ Lie algebra brackets, under the same conditions, we get again $\bR= U(1)$ and an Aloff-Wallach space \cite{AlWa75}  $\cV = SU(3)/U(1)$ 
is reductive in the split structure. We stress once more that the above is a consequence of the requirement of the existence of {\em canonical} covariant gauge-natural conserved quantities.

The calculations above can be applied to  the Lie algebra of the structure group of the $(1,1)$-gauge-natural bundle of principal connections 
$W_{4}^{(1,1)}SU(3)= T^1_4 SU(3)\rtimes GL(4, \mathbb{R})
\simeq (SU(3) \ltimes (\mathbb{R}^{4})^{*} \ten \, \mathfrak{su}(3)) \rtimes  GL(4, \mathbb{R}) 
\simeq 
[(\mathbb{R}^{4})^{*} \ten \, \mathfrak{su}(3) \rtimes  GL(4, \mathbb{R})] \rtimes SU(3)$. 

Indeed, let us specialize to vertical vector fields on the bundle of connections {\em which are gauge-natural lifts}, \ie (according with \cite{FaFr03} p. $95$) for $\hat{\Xi}^Z_\alpha= d_\alpha\Xi^Z + c^{Z}_{LM}\Xi^L \ome^{M}_\alpha$ , where $\Xi^Z (x) \mathfrak{b}_Z = \Xi^L (x) (TR_g)^L_Z \der_L$ is an infinitesimal gauge automorphism of the underlying  $SU(3)$ principal bundle. 
We see that only the Lie algebra $\mathfrak{su}(3)$ play a r\^ole in the expressions of the gauge-natural lift $\hat{\Xi}^Z_\alpha$; we can therefore still apply the above equations (obtained for simplicity in the case of a semi-simple group) and obtain that $\cV = W_{4}^{(1,1)}SU(3)/U(1)$ is reductive in the split structure.

In particular, for any vertical lift, $(\cL_{\hat{\Xi}}\ome)^A_\mu= -d_\mu \Xi^A - c^A_{BC}\Xi^B\ome^C_\mu = - (\hat{\Xi}_V)^A_\mu$, we see that, as expected,
$(\hat{\Xi}_V)^A_\mu = \hat{\Xi}^A_\mu = \hat{\nabla}_\mu \Xi^A$, \ie the vertical part of a gauge-natural lift of a {\em vertical} vector field coincides with the gauge-natural lift itself and equals a suitably defined covariant derivative of  $\Xi^Z (x)$. 
Therefore, it is now clear that also the Lie derivative of fields is constrained (a fact pointed out in \cite{PaWi08,PaWi13}).
Let us then consider vertical tangent vector fields which are fundamental vector fields; in this case $\Xi^Z $ have to be constants and we have that $d_\alpha\Xi^Z = 0$. Being in this case $\Xi^Z_\mu= c^{Z}_{LM}\Xi^L \ome^{M}_\mu$,
the above implies that $\ome^{M}_\mu$ is constrained (see also  \cite{FFPW08,FFPW10}).

Note that the results obtained in the present example, in principle, could be extended to a Yang--Mills theory on a generic metric space-time, the restriction to a Minkowskian background being here mainly motivated by the fact that calculations are simplified. Nonetheless, already at this simple level they provide physically important consequences; indeed the relation with confinement phases in non-abelian gauge theories \cite{'tHo81} deserves further study. As for the interest in Physics, it is also worth to mention the possibility to extend the concept of a  Higgs field defined here to principal superbundles in the category of $G$-supermanifolds; see in particular \cite{Sar08}.

\section*{Acknowledgements} 
Research supported by Department of Mathematics - University of Torino under research project {\em Algebraic and geometric structures in mathematical physics and applications (2016--2017)}  (MP) and written under the auspices of GNSAGA-INdAM.


\end{document}